\documentclass[12pt]{article}

\usepackage{amsmath}
\usepackage{amsfonts}
\usepackage{latexsym}
\usepackage{mathrsfs}
\usepackage{graphicx}

\renewcommand{\atop}{, }

\newcommand{\N}{{\mathscr N}}

\newcommand{\F}{{\mathscr F}}

\def\S{{\mathscr S}}

\def\e{{\rm e}}

\def\l{\langle}
\def\r{\rangle}
\def\pr{\partial}

\newcommand{\BRS}{{\rm BRS}}

\newcommand{\half}{{\textstyle \frac{1}{2}}}
\def\quar{{\textstyle \frac{1}{4}}}

\newcommand{\Geff}{\Gamma_{\rm eff}}
\newcommand{\Gcl}{\Gamma_{\rm cl}}

\newcommand{\dx}{\!\!{\rm d}^4x\,}
\newcommand{\dv}{\!\!{\rm d}^3x\,}
\newcommand{\dz}{\!\!{\rm d}^4z\,}
\newcommand{\dt}{\!\!{\rm d}t\,}
\newcommand{\dnx}[2]{\!\!{\rm d}^{#1}{#2}\,}

\newcommand{\al}{\alpha}
\def\da{{\dot\alpha}}
\def\be{\beta}
\def\db{{\dot\beta}}

\newcommand{\eps}{\varepsilon}

\newcommand{\tfr}[2]{{\textstyle \frac{#1}{#2}}}

\newcommand{\fdq}[2]{\frac{\delta #1}{\delta #2}}

\newcommand{\pdq}[2]{\frac{\partial #1}{\partial #2}}

\renewcommand{\i}{{\rm i}}

\newcommand{\op}{{\rm Op}}
\newcommand{\brs}{{\rm BRS}}
\newcommand{\T}{{\rm T}}

\newcommand{\s}{\mbox{\bf s}\,}

\newcommand{\out}{{\rm out}}
\renewcommand{\in}{{\rm in }}

\begin{document}

\thispagestyle{empty}
\vspace*{-5mm}
\begin{flushright}
hep-th/0101153\\  October 2000
\end{flushright}

\vspace{0.5cm}
\begin{center}
{\Large \bf Supersymmetry Transformation of Quantum Fields}

\vspace{0.5cm}

%{\parindent0cm
Christian Rupp \footnote{Email: Christian.Rupp@itp.uni-leipzig.de}\\
Rainer Scharf \footnote{Email: Rainer.Scharf@itp.uni-leipzig.de}\\
Klaus Sibold \footnote{Email: Klaus.Sibold@itp.uni-leipzig.de}
%}
\vspace{0.3cm}

Institut f{\"u}r Theoretische Physik\\
Universit{\"a}t Leipzig\\
Augustusplatz 10/11\\
D - 04109 Leipzig\\
Germany
\end{center}

\vspace{1cm}

\begin{center}
\parbox{12cm}{
\centerline{\small \bf Abstract} 
{ \small \noindent 
In the Wess-Zumino gauge, supersymmetry transformations become
non-linear and are usually incorporated together with BRS
transformations in the form of Slavnov-Taylor identities, such that
they appear at first sight to be even non-local. Furthermore, the
gauge fixing term breaks supersymmetry.
In the present paper, we clarify in which sense supersymmetry is still
a symmetry of the system and how it is
realized on the level of quantum fields.
 }}
\end{center}

\vspace*{15mm}
\begin{tabbing}
PACS numbers: \= 03.70.+k, 11.15.Bt, 11.30.Pb\\
 Keywords:\>
Quantum Field Theory, Supersymmetry, Wess-Zumino Gauge.
\end{tabbing}
\newpage

\section{Introduction}
Electroweak processes in todays particle physics are very successfully
described by the standard model \cite{GSW}. The precision of the LEP II
experiments will require two loop corrections \cite{elweak}, which in
turn will be calculable because the algebraic method of
renormalization has been 
extended to the standard model \cite{EKhabil}.
This method permits an unabiguous construction of the model via the
Slavnov-Taylor (ST) identitiy and a rigid and a local gauge Ward identity
(WI) in a way independent from the applied renormalization scheme.
At present the only viable alternative to the SM is a supersymmetric
extension thereof, e.g.\ a minimal one (MSSM). Here the status of loop
corrections is much less satisfactory because the effect of the
renormalized supersymmetry WI has not yet been systematically taken into
account. A first step into this direction has been undertaken in
\cite{HKS} for the susy extension of QED. It applies techniques developped in
\cite{White, MPW} and gives clear prescriptions on how to
establish and use the susy WI.
This is non-trivial because in the chosen Wess-Zumino gauge \cite{WZgauge} the
susy transformations are non-linear and -- as is shown in \cite{HKS} --
also non-local.
The question then arises in which sense susy is still a symmetry of the
system. 
Clearly, it is realized on the Green functions but is it realized on the
local fields, on the state space? Does there exist a susy charge which
commutes with the S-matrix?
In the present paper we wish to answer these questions for the
Wess-Zumino model and then for SQED, basing our
analysis on the results of \cite{HKS}.

\section{The Wess-Zumino model without auxiliary fields
  \label{sec:wess-zumino-model}}
\setcounter{equation}{0}

The model comprises a complex scalar and a (Weyl) spinor field together with their conjugates, forming the classical action

\begin{align}
\Gamma_{\rm WZ} = \int \dx \Bigl\{&
\pr^\mu \phi \pr_\mu \bar \phi - m^2 \phi \bar\phi
+ \tfr{\i}{2} \psi \sigma^\mu \pr_\mu \bar \psi + \quar m ( \psi\psi +
\bar\psi \bar\psi) \nonumber \\
&
- \tfr{1}{16} g^2 \phi^2 \bar \phi^2 + \tfr{g}{8} \psi\psi\phi +
\tfr{g}{8} \bar \psi \bar \psi \bar \phi 
  -\quar mg \phi\bar\phi (\phi+\bar\phi) \Bigr\}
\label{eq:WZaction}
\end{align}

It has one common mass for all fields and only one coupling. This is due to supersymmetry and parity which leave invariant this action.
Without auxiliary fields the susy transormations are non-linear and their algebra closes only on the mass shell - a problem for renormalization. It has been solved in \cite{PSWZ} by introducing external fields coupled to the non-linear variations and formulating the Ward identity (WI) for susy as a $\Gamma$-bilinear equation. Crucial is the fact that in $\Geff$ appears an expression bilinear in the external fields. Meanwhile this method has been greatly refined \cite{White, MPW} and systematized, hence we shall present the relevant results in this form.

\subsection{BRS transformations, ST identity} \label{sec:brs-transf-st}
The systematic procedure for treating susy without auxiliary fields consists
in promoting the parameters of the transformations to constant ghosts. Those
carry a Grassmann number which is always opposite to its statistics: the
ghosts of susy ($\epsilon^\al$, $\bar \epsilon_\da$) commute, the translations ($\omega_\mu$) anticommute.
In order to have nilpotent transformations the ghosts too have to transform.
With the assignment
\begin{align}
\s \phi &= \eps^\al \psi_\al -\i \omega^\mu \pr_\mu \phi  \\
\s \psi^\be &= \tfr{g}{2} \eps^\be \bar \phi^2 +2 \eps^\be m \bar \phi
-2\i \bar \eps_\da
\sigma^{\mu\be\da} \pr_\mu \phi -\i \omega^\mu \pr_\mu \psi^\be \\
\s \bar \phi &= -\bar \eps_\da \bar \psi^\da -\i \omega^\mu \pr_\mu \bar \phi
\\
\s \bar \psi^\db &= -\tfr{g}{2} \bar \eps^\db \phi^2 -2 \eps^\db m \phi
- 2\i \eps_\al
\sigma^\mu{}^\al{}^\db \pr_\mu \bar \phi -\i \omega^\mu \pr_\mu \bar
\psi^\da \\
\s \eps^\al &=0 = s \bar \eps^\da\\
\s \omega^\mu &= 2 \eps \sigma^\mu \bar \eps
\end{align}
one finds
\begin{align}
\s^2 \phi &=0=\s^2\bar \phi 
\label{eq:WZnil1}\\
\s^2 \psi^\be &= - 4 \eps^\be \bar \eps^\db \fdq{\Gamma_{\rm WZ}}{\bar
  \psi^\db} 
\label{eq:WZnil2}\\
\s^2 \bar \psi^\db &= 4 \eps^\be \bar \eps^\db \fdq{\Gamma_{\rm
    WZ}}{\psi^\be}
\label{eq:WZnil3}
\end{align}
In order to obtain off-shell closure one introduces external fields
coupled to the non-linear variations and a term bilinear in these
external fields
\begin{align}
\Gamma_{\rm ext} &= \int \left\{ Y^\al \, s \psi_\al + \bar Y_\da \, s \bar
  \psi^\da + 4 \eps^\al Y_\al\, \bar \eps_\da \bar Y^\da \right\} \label{eq:GammaWZext}\\
\Gamma_{\rm cl} &= \Gamma_{\rm WZ} + \Gamma_{\rm ext}
\end{align}
The symmetry of the model can now be expressed as a ST identity
\begin{equation}
\S(\Gamma) \equiv \int \left\{ s\phi \fdq{\Gamma}{\phi} + s \bar \phi
  \fdq{\Gamma}{\bar \phi} + \fdq{\Gamma}{Y_\al} \fdq{\Gamma}{\psi^\al} +
  \fdq{\Gamma}{{\bar Y}^\da} \fdq{\Gamma}{\bar \psi_\da} \right\} +
  s\omega^\mu \pdq{\Gamma}{\omega^\mu} \,\,=\, 0\,.
\label{eq:WZSTI}
\end{equation}
It holds for $\Gamma=\Gcl$. One observes furthermore that the linearized operator
\begin{align}
\S_\Gamma &\equiv \int \left( \s \phi \fdq{}{\phi} + \s\bar\phi\fdq{}{\bar\phi} + \fdq{\Gamma}{Y_\al}\fdq{}{\psi^\al} + \fdq{\Gamma}{\psi^\al}\fdq{}{Y_\al} + \fdq{\Gamma}{\bar Y^\da}\fdq{}{\bar Y_\da} + \fdq{\Gamma}{\bar Y^\da} \fdq{}{\bar Y_\da} \right) + \s 
\omega^\mu \pdq{}{\omega^\mu}
\end{align}
satisfies
\begin{align}
\S_\Gamma^2 &=0 \\
\S_\Gamma \s \phi &= 0 \label{sGammasPhi}\\
\S_\Gamma \S(\Gamma) &=0
\end{align}
for $\Gamma = \Gcl$.
Crucial for the validity of ($\S_\Gamma^2=0$) is the presence of the $Y
\bar Y$-term in (\ref{eq:GammaWZext}). 
It clearly contributes those eq. of motion terms which guarantee on the
functional level nilpotency as opposed to (\ref{eq:WZnil2}),
(\ref{eq:WZnil3}) where -- on the level of elementary fields -- it does not
hold. 
Eqn. (\ref{sGammasPhi}) will serve as a consistency condition for constraining
potential non-symmetric higher order corrections to the ST identity
(\ref{eq:WZSTI}). 

The renormalization is now straightforward \cite{PSWZ} and yields the ST
identity (\ref{eq:WZSTI}) if one imposes in addition 
\begin{equation}
\pdq{\Gamma}{\omega^\mu} = \pdq{\Gcl}{\omega^\mu} \label{eq:WZnorm1}
\end{equation}
which is possible, and recursively
\begin{equation}
(\S_\Gamma \s \phi) =0\,. \label{eq:WZnorm2}
\end{equation}
(\ref{eq:WZnorm1}) and (\ref{eq:WZnorm2}) imply
\begin{align}
\pdq{\Gamma}{\omega^\mu} &= -\i \int \left( Y^\al \pr_\mu \psi_\al + \bar Y_\da \pr_\mu \bar \psi^\da \right) \label{eq:WZnorm1a}\\
\eps^\al \pdq{\Gamma}{Y^\al} &= -\i \eps^\al \omega^\mu \pr_\mu \psi_\al +2\i
\eps \sigma^\mu \bar \eps \pr_\mu \phi
\label{eq:WZnorm2a}
\end{align}
In order to prepare for the subsequent discussion we go via Legendre transformation over to the connected Green functions
\begin{align}
-j &= \fdq{\Gamma}{\phi}\,, \qquad \fdq{Z_c}{Y^\al} = \fdq{\Gamma}{Y^\al}\,, \qquad -\eta_\al = \fdq{\Gamma}{\psi^\al} \\
Z_c &= \Gamma + \int \left( j \phi + \bar j \bar \phi + \eta_\al
  \psi^\al + \bar \eta _\da \bar \psi^\da \right)
\end{align}
and then to the general Green functions
\begin{equation}
Z=\e^{\i Z_c}
\end{equation}
The validity of the ST identity (\ref{eq:WZSTI}) entails the existence of
a conserved current for a {\em local} generalized BRS transformation 
\begin{equation}
\S_{\rm loc}(y) Z = [\pr^\mu J_\mu(y)]\cdot Z
\label{WZSTIloc}
\end{equation}
where
\begin{align}
\S_{\rm loc} & \equiv \i j \eps^\al \fdq{}{\eta^\al} - \omega^\mu
\pr_\mu j \fdq{}{j} -\i \bar j \bar \eps_\da \fdq{}{\bar\eta_\da} -
\omega^\mu \pr_\mu  \bar j \fdq{}{\bar j}
-\i \eta^\al \fdq{}{Y^\al} -\i \bar \eta_\da \fdq{}{\bar Y_\da}
\nonumber \\
&\quad -2 \eps\sigma^\mu \bar \eps \left( Y^\al \pr_\mu \fdq{}{\eta^\al}
  + \bar Y_\da \pr_\mu \fdq{}{\bar \eta_\da} \right) \,.
\end{align}

\subsection{Supersymmetry transformations of functionals}\label{sec:supersymm-transf-fun}
The presence of constant ghosts in the theory permits the subdivision of functionals and sometimes also of their transformations into sectors of fixed ghost number -- this is commonly called a ``filtration'' \cite{White, MPW}.
We expand 
\begin{equation}
\Gamma = \Gamma^{(0)} + \Gamma^{(1)} + \dots
\end{equation}
according to the number of constant ghosts.
Inserting this expansion into the ST identity the latter can first be
rewritten as 
\begin{equation}
\S(\Gamma) = \S(\Gamma^{(0)}) + \S_{\Gamma^{(0)}} \Gamma^{(1)} +
O(\epsilon^2) \label{eq:1}
\end{equation}
where 
\begin{equation}
\S(\Gamma^{(0)}) = \int \left( \s\phi \fdq{\Gamma^{(0)}}{\phi} +
  \fdq{\Gamma^{(0)}}{Y} \fdq{\Gamma^{(0)}}{\psi} \right) + c.c. \label{eq:2}
\end{equation}
(\ref{eq:1}) is verified by observing that $\fdq{\Gamma^{(0)}}{Y}=0$.
Using
\begin{equation}
\S_{\Gamma^{(0)}} \Gamma^{(1)} = \int \left( \fdq{\Gamma^{(1)}}{Y} \fdq{\Gamma^{(0)}}{\psi} + c.c. \right) + \s\omega^\mu \pdq{\Gamma}{\omega^\mu} + O(\epsilon^2)
\end{equation}
one can write
\begin{equation}
\S(\Gamma) = (\S_\Gamma)^{(1)} \Gamma^{(0)} + O(\epsilon^2)\,,
\end{equation}
hence, since the ST identity holds order by order in the ghosts
\begin{equation}
\S_\Gamma^{(1)} \Gamma^{(0)} =0\,.
\end{equation}
We decompose
\begin{align}
\S_\Gamma^{(1)} &= \epsilon^\al W_\al + \bar \epsilon^\da \bar W_\da +
\omega^\mu W_\mu - 2 \epsilon \sigma^\mu \bar \epsilon \pdq{}{\omega^\mu}
\label{eq:3}
\end{align}
such that functional differential operators are defined by differentiation
w.r.t.\ the ghosts
\begin{align}
W_\al &= \pdq{}{\epsilon^\al} \S_\Gamma^{(1)} \Bigr|_{\epsilon=\omega=0}
\label{eq:4} \\
\bar W_\da &= \pdq{}{\bar \epsilon^\da} \S_\Gamma^{(1)} \Bigr|_{\epsilon=\omega=0}
\label{eq:5} \\
W_\mu &= \pdq{}{\omega^\mu} \S_\Gamma^{(1)} \Bigr|_{\epsilon=\omega=0}
\label{eq:6}
\end{align}
The filtration of $\S_\Gamma$ contains besides $\S_\Gamma^{(1)}$
\begin{align}
\S_\Gamma^{(0)} &= \fdq{\Gamma^{(0)}}{\psi^\al} \fdq{}{Y_\al} +
  c.c. \label{eq:7} \\
\S_\Gamma^{(2)} &= \fdq{\Gamma^{(2)}}{Y_\al} \fdq{}{\psi_\al} +
  c.c. \label{eq:8} 
\end{align}
Next we exploit the fact that $\S_\Gamma$ is nilpotent on functionals
$\F$ for which
\begin{equation}
\eps^\al \pdq{\F}{Y^\al} = -\i \eps^\al \omega^\mu \pr_\mu \psi_\al +2\i
\eps \sigma^\mu \bar \eps \pr_\mu \phi
\end{equation}
holds.
This yields
\begin{align}
\S_\Gamma^{(0)} \S_\Gamma^{(0)} &=0 \label{eq:filtnilpot0} \\
\{ \S_\Gamma^{(0)}, \S_\Gamma^{(1)}\} &=0 \label{eq:filtnilpot1} \\
\S_\Gamma^{(1)} \S_\Gamma^{(1)} + \{ \S_\Gamma^{(0)}, \S_\Gamma^{(2)}\}
&=0 \label{eq:filtnilpot2}
\end{align}
Projecting out of $\S_\Gamma^{(1)}\S_\Gamma^{(1)}$ the
$\epsilon\bar\epsilon$-part one finds
\begin{align}
\S_\Gamma^{(1)}\S_\Gamma^{(1)} \Bigr|_{\epsilon\bar\epsilon} &=
\epsilon^\al \bar \epsilon^\da \left( \{ W_\al, \bar W_\da \} - 2
  \sigma^\mu_{\al\da} W_\mu \right)\,. \label{SUSYalgebrafunctional}
\end{align}
(\ref{eq:filtnilpot2}) with (\ref{SUSYalgebrafunctional}) shows that even on
$Y$-independent functionals $\F$, the SUSY algebra closes only up to equation
of motions terms $\fdq{\F}{\psi}$.

%------------------------------------------------------
 \subsection{Symmetry transformations on quantum fields}
 \label{sec:symm-transf-quant} 
%------------------------------------------------------

Vertex functions can essentially be understood as matrix elements of
operators but are certainly not the ideal tool for revealing the underlying properties of the respective operators, hence we study general Green functions.

As the simplest example we identify the energy-momentum operator and its action on field operators.
We therefore insert (\ref{eq:WZnorm2a}) into (\ref{eq:WZSTI}) and permit
$\omega_\mu$, $\eps^\al$ and $\bar \eps_\da$ to be local,
$\omega_\mu=\omega_\mu(x)$ etc., 
\begin{equation}
\S_{\rm loc} Z = [\pr^\mu J_\mu + \pr^\mu \eps^\al K_{\mu\al} +
\pr^\mu \bar \eps^\da \bar K_{\mu\da} + \pr^\mu \omega^\nu K_{\mu\nu}
] \cdot Z\,.
\label{WZSTIlocloc}
\end{equation}
Differentiating
w.r.t.\ $\omega_\mu(z)$ and integrating over $z$ yields the local WI 
\begin{align}
\left. w_\mu \Gamma\right|_{\epsilon=\bar\epsilon=0 \atop Y=\bar Y=0} &\equiv \i \left. \left( \pr_\mu \phi \fdq{}{\phi} + \pr_\mu \bar \phi \fdq{}{\bar \phi} + \pr_\mu \psi \fdq{}{\psi} + \pr_\mu \bar \psi \fdq{}{\bar \psi} \right) \Gamma \right|_{\genfrac{}{}{0pt}{}{\epsilon=\bar\epsilon=0}{Y=\bar Y=0}} \\
&= \left. \left[\pr^\nu T_{\nu\mu} \right] \cdot \Gamma
\right|_{\genfrac{}{}{0pt}{}{\epsilon=\bar\epsilon=0}{Y=\bar Y=0}} \,,
\end{align}
with $T_{\nu\mu}=\pr_{\omega^\nu} J_\mu$.
(~(\ref{eq:WZnorm1a}) taken at $Y=0=\bar Y$ guarantees that no other contributions depending on $\omega_\mu$ show up.)
Translated onto $Z$ we obtain
\begin{equation}
\left( \pr_\mu j_\phi \fdq{}{j_\phi} + \pr_\mu \eta \fdq{}{\eta} +
  c.c. \right) Z \Bigr|_{Y=0} = -\i [\pr^\nu T_{\nu\mu}]\cdot Z
\label{eq:transWIZ}
\end{equation}
a local WI for the translations, with $T_{\nu\mu}$ being the energy-momentum tensor.
By differentiating with respect to a suitable combination of sources we
generate on the right hand side the Green function $-\pr^\nu \l \T
(T_{\nu\mu} X) \r$, where $X$ stands for an arbitrary number of
elementary fields; on the left hand side there will always occur Green
functions with one field argument less and a $\delta$-function
instead. Multiplying with inverse propagators and going on mass shell we
obtain zero on the l.h.s.\ (as a consequence of the $\delta$-functions) and
on the r.h.s.\ the respective matrix elements of the operator
$\pr^\nu T_{\nu\mu}^\op$. Hence
\begin{equation}
\pr^\nu T_{\nu\mu}^\op =0\,,
\label{momentumconservation}
\end{equation}
the energy-momentum tensor is conserved.

Defining the energy-momentum operator by
\begin{equation}
P_\nu^\op = \int \dv T_{0\nu}^\op\,,
\end{equation}
(\ref{momentumconservation}) means that $P_\nu^\op$ is time
independent. This detailed presentation served as an illustration of the LSZ reduction technique which permits one to generate operator relations out of Green functions \cite{Kugo}.

In order to obtain the transformation law of field operators we deduce from
(\ref{eq:transWIZ}) 
\begin{equation}
\pr_\nu^y \delta(y-x) \, \l \varphi(y) \, X \r = -\i \pr^{y\mu} \l \T
(T_{\mu\nu}(y) \varphi(x) X ) \r\,.
\end{equation}
Reduction yields
\begin{equation}
\pr_\nu^y \delta(y-x) \varphi^\op(y) = -\i \pr^{y\mu} \T (T_{\mu\nu}(y) \varphi(x))^\op\,.
\end{equation}
Integration over $\int \!\!{\rm d}^3y$ and $\int_{x^0-\eps}^{x^0+\eps} {\rm d}y^0$ leads to
\begin{equation}
\pr_\mu \varphi^\op(x) = \i [P_\mu, \varphi(x)]^\op\,,
\qquad \text{for } \varphi=\phi, \bar\phi, \psi, \bar\psi
\label{eq:transfield}
\end{equation}
the well-known transformation law of a field operator under translations.

We learn from this result that the local version of the ST identity on $Z$
(\ref{WZSTIlocloc}) should also be most useful for deriving the susy current,
charge and field transformation. 
\begin{multline}
\i \delta(y-z) j(z) \fdq{Z}{\eta_\al(z)} 
+ \i \eta^\be \left[ \fdq{\Geff}{\epsilon^\al}\right] \cdot \left[\fdq{\Geff}{Y^\be}\right] \cdot Z +  \eta^\be \left[ \fdq{^2\Geff}{\epsilon^\al(z)\, \delta Y^\be(y)} \right] \cdot Z 
 \\
+ \i\bar \eta^\db \left[ \fdq{\Geff}{\epsilon^\al}\right] \cdot
\left[\fdq{\Geff}{\bar Y^\db}\right] \cdot Z +  \eta^\be \left[
  \fdq{^2\Geff}{\epsilon^\al(z)\, \delta \bar Y^\db(y)} \right] \cdot
Z
 \\
 =
\left[ \pr^\mu_y \fdq{J_{\mu}(y)}{\eps^\al(z)} \right] \cdot Z 
  + \i [\pr^\mu J_\mu(y)]\cdot \left[ \fdq{\Geff}{\eps^\al(z)}
 \right] \cdot Z
  + \pr^\mu_y \delta(y-z) \left[ K_{\mu\al}(y) \right]
\cdot Z
\label{eq:WZsusyloc}
\end{multline}
Since $\eps$ has ghost number one, it may occur only in combination
with $Y$, hence no $\epsilon$-dependent term survives
at $Y=0$, and the double insertions in (\ref{eq:WZsusyloc}) are
absent. 
One can safely integrate over $z$ which yields
\begin{align}
\i j(y) \fdq{Z}{\eta_\al(y)} - \eta^\be(y)
\fdq{\pr_{\epsilon^\al}\Geff}{Y^\be(y)} + \i \bar \eta^\db
(y)\fdq{\pr_{\epsilon^\al}\Geff}{\bar Y^\db(y)} &= [\pr^\mu
J_{\mu\al}(y)]\cdot Z \,,
\end{align}
where $J_{\mu\al} \equiv \pr_{\eps^\al} J_\mu$  represents the susy
current. 
From here on one may just perform all steps as before for the translations and obtain 
\begin{equation}
\pr^\mu J_{\mu\al}^\op =0\,,
\end{equation}
the current conservation.
With the charge
\begin{equation}
Q_\al \equiv - \int \dv J_{0\al}(x)
\end{equation}
the field transformations follow,
\begin{align}
\i [Q_\al, \phi]^\op &= \psi_\al^\op \label{eq:WZsusyfield1}\\
\i \{ Q_\al, \psi_\be\}^\op &= \left(\fdq{}{Y^\be}\pr_{\epsilon^\al}\Geff
\right)^\op \equiv \delta_\al \psi_\be^\op\,. \label{eq:WZsusyfield2}
\end{align}
Analogously for the conjugate quantities.

As a consequence of (\ref{eq:WZsusyfield1}), (\ref{eq:WZsusyfield2}) and
the translations one derives the validity of the algebra 
\begin{align}
\left[ \{ \i Q_\al, \i \bar Q_\da \}, \phi \right] &= \{\i Q_\al, [\i \bar Q_\da, \phi ]\} + \{ \i \bar Q_\da, [\i Q_\al, \phi ]\} \nonumber \\
&= \{\i \bar Q_\da, \psi_\al \} = \fdq{}{Y^\al}\pr_{\bar \eps^\da} \Geff = 2\i \sigma^\mu_{\al\da} \pr_\mu \phi \nonumber \\
&= -2 \sigma^\mu_{\al\da} [P_\mu, \phi]
\end{align}
Similarly,
\begin{align}
\left[ \{ \i Q_\al, \i \bar Q_\da \}, \psi_\be \right] &=
\left[ \i Q_\al, \fdq{}{Y^\be}\pr_{\bar \eps^\da} \Geff \right]
+ \left[ \i \bar Q_\da, \fdq{}{Y^\be}\pr_{\eps^\al} \Geff \right]
\nonumber\\
&= -2\i \sigma^\mu_{\al\da} \pr_\mu \psi_\be \,,
\end{align}
where the second equality is established by differentiating
(\ref{WZSTIlocloc}) w.r.t.\ $\eps^\al$, $\bar\eps^\da$, $Y^\be$ and
performing the LSZ reduction. Thus we have
\begin{equation}
\{ Q_\al, \bar Q_\da \} = 2 \sigma^\mu_{\al\da} P_\mu \,.
\end{equation}
The current conservation implies that $Q_\al$, $\bar Q_\da$ act as symmetry
operators on the Hilbert space of the theory, with the transformation of
field variables given by (\ref{eq:WZsusyfield1}), (\ref{eq:WZsusyfield2}). 
This field transformation law is non-linear (for $\psi$), but it is local!
And in this sense one can say that the transformations as given on the functionals (\ref{SUSYalgebrafunctional}) resemble the operator law here and hint indeed correctly to a local symmetry.
We obtain a closed algebra here because the field operators obey the equations
of motion. 

This concludes our discussion of the Wess-Zumino model.

%---------------------------------------------------
\section{SQED in the Wess-Zumino gauge} \label{sec:sqed-wess-zumino}
\setcounter{equation}{0}

%---------------------------------------------------
The susy and gauge invariant action of SQED in the Wess-Zumino gauge reads
as follows
\begin{align}
\Gamma_{\rm SQED} &= -\quar \int F_{\mu\nu}F^{\mu\nu} + \half \int \bar
\gamma \, ( \i \gamma^\mu)\,  \pr_\mu \gamma \nonumber \\
& \quad + \int |D_\mu \phi_L|^2 + \int |D_\mu\phi_R^\dagger |^2 + \bar \Psi
\, (\i \gamma^\mu)\,  D_\mu \Psi \nonumber \\
& \quad - \sqrt{2} e Q_L \int \left( \bar \Psi P_R \gamma \phi_L - \bar
  \Psi P_L \gamma \phi_R^\dagger + \phi_L^\dagger \bar \gamma P_L \Psi -
  \phi_R \bar \gamma P_R \Psi \right) \nonumber \\
& \quad - \half \int \left(  e Q_L  |\phi_L|^2 + e Q_R |\phi_R|^2\right)^2
\nonumber \\
& \quad -m \int \bar \Psi \Psi - m^2 \int \left( |\phi_L|^2 +
  |\phi_R|^2\right)\\ 
\intertext{where}
D_\mu &\equiv \pr_\mu + \i e Q A_\mu \\
F_{\mu\nu} &\equiv \pr_\mu A_\nu - \pr_\nu A_\mu
\label{eq:GammaSQED}
\end{align}
and $(A^\mu, \lambda^\al, \bar \lambda_\da)$ denote the photon and
Weyl-photino resp., whereas $(\psi_L^\al, \phi_L)$ and $(\psi^\al_R,
\phi_R)$ refer to two chiral multiplets with charges $Q_L=-1$, $Q_R=+1$.
The electron Dirac spinor and the photino Majorana spinor are given by
\begin{equation}
\Psi = \binom{\psi_L^\al}{\bar \psi_R^\da}\,, \qquad
\gamma = \binom{-\i \lambda_\al}{\phantom{-}\i \bar\lambda^\da}\,.
\end{equation}
In this gauge model susy transformations are non-linear not only because the auxiliary fields have been eliminated but also because longitudinal components of the vector superfield are being transformed away. This causes an additional problem:
every susy transformation has to be followed by a field dependent gauge transformation such that one stays in the Wess-Zumino gauge.
Hence only gauge invariant terms can at the same time be supersymmetric; a gauge fixing term can never be susy invariant.
The task is therefore to construct the model and then physical (i.e.\ gauge invariant) quantities ``modulo gauge fixing''.

%---------------------------------------------------
\subsection{BRS transformations, ST identity} \label{sec:brs-transf-st-1}
%---------------------------------------------------
The problems that susy transformations are non-linear, close only on-shell
and that gauge fixing is not supersymmetric have all been overcome by going
to a BRS formulation of all transformations and introducing suitable
external fields, in particular also terms in $\Geff$ which are bilinear in
the latter \cite{White, MPW, HKS}.
These generalized BRS transformations have the form
\begin{subequations}
\begin{align}
\s A_\mu &=  \partial_\mu c + \i\eps\sigma_\mu\bar\lambda 
              -\i \lambda\sigma_\mu\bar\eps -\i\omega^\nu\partial_\nu
              A_\mu \, ,\\
\s\lambda^\alpha & =  \phantom{-} \tfr{\i}{2} (\eps\sigma^{\rho\sigma})^\alpha 
             F_{\rho\sigma} - \i\eps^\alpha\,
             eQ_L(|\phi_L|^2-|\phi_R|^2) -\i\omega^\nu\partial_\nu  
             \lambda^\alpha \, ,\\
\s\bar\lambda_\da & =  -\tfr{\i}{2} (\bar \eps\bar
             \sigma^{\rho\sigma}) 
             _\da F_{\rho\sigma} - \i\bar\eps_\da\,  
             eQ_L(|\phi_L|^2-|\phi_R|^2)  
             -\i\omega^\nu\partial_\nu \bar\lambda_\da  \, ,\\
\s\phi_L & =  -\i eQ_L c\,\phi_L +\sqrt{2}\, \eps\psi_L -
             \i\omega^\nu\partial_\nu \phi_L \, ,\\
\s\phi_L^\dagger & =  \phantom{-} \i eQ_L c\,\phi_L^\dagger +\sqrt{2}\, 
           \bar\psi_L\bar\eps - \i\omega^\nu\partial_\nu \phi_L^\dagger
           \ ,\\  
\s\psi_L^\alpha & =  -\i eQ_L c\,\psi_L^\alpha - \sqrt{2}\, 
         \eps^\alpha\, m\phi_R^\dagger 
         -\sqrt{2}\, \i (\bar\eps\bar\sigma^\mu)^\alpha D_\mu\phi_L 
         -\i\omega^\nu\partial_\nu \psi_L^\alpha
\ ,\\
\s \bar\psi_{L\da} & =  \phantom{-} \i eQ_L c\,\bar\psi_{L\da}
         + \sqrt{2}\,\bar\eps_\da\, m\phi_R + \sqrt{2}\, \i
         (\eps\sigma^\mu)_\da (D_\mu\phi_L)^\dagger 
         -\i\omega^\nu\partial_\nu \bar\psi_{L\da}
\ ,\\
\s c & =  2\i\eps\sigma^\nu\bar\eps A_\nu -\i\omega^\nu\partial_\nu c
\ ,\\
\s\eps^\alpha & =  0
\ ,\\
\s\bar\eps^\da & = 0
\ ,\\
\s\omega^\nu & =  2\eps\sigma^\nu\bar\eps
\ ,\\
\s\bar c & =  B - \i\omega^\nu\partial_\nu \bar c
\ ,\\
\s B & =  2\i\eps\sigma^\nu\bar\eps \partial_\nu \bar c 
         -\i\omega^\nu\partial_\nu B
\end{align}
\label{BRStrans}
\end{subequations}
A suitable form of gauge fixing turns out to be
\begin{align}
\Gamma_{\rm g.f.} &= \int \s \left( \bar c \pr A + \tfr{\xi}{2} \bar c
  B \right)
\\
& = \int \Bigl( B \pr A + \tfr{\xi}{2} B^2 - \bar c \Box c - \bar c \pr^\mu \left( \i \eps \sigma_\mu \bar \lambda - \i
  \lambda \sigma_\mu \bar \eps \right) + \xi \i \eps \sigma^\nu \bar \eps
\pr_\nu \bar c \bar c \Bigr)
\label{Gammagf}
\end{align}
The non-linear transformations will be defined via their coupling to
external fields
\begin{align}
\Gamma_{\rm ext} &= \int \left( Y^\al_\lambda \s\lambda_\al + Y_{\bar
    \lambda \da} \s \bar\lambda^\da + Y_{\phi_L} \s\phi_L +
    Y_{\phi_L^\dagger} \s\phi_L^\dagger + Y_{\psi_L}^\al \s \psi_{L\al} +
    Y_{\bar\psi_L\da} \s\bar\psi_L^\da + (L \to R) \right)
\end{align}
complemented by well specified correction terms in higher orders.

The contributions which are bilinear in the external fields have the form
\begin{align}
\Gamma_{\rm bil} &= - \int \left( (Y_\lambda \eps) (\bar \eps
  Y_{\bar\lambda}) + 2 (Y_{\psi_L} \eps) (\bar \eps Y_{\bar\psi_L}) + 2
  (Y_{\psi_R}\eps)(\bar\eps Y_{\bar\psi_R})\right)\,.
\end{align}
The classical action 
\begin{align}
\Gamma_{\rm cl} &= \Gamma_{\rm SQED} + \Gamma_{\rm g.f.} + \Gamma_{\rm ext}
+ \Gamma_{\rm bil}
\end{align}
satisfies then a Slavnov-Taylor identity which can be extended \cite{HKS}
to all orders in the loop expansion for the vertex functional $\Gamma$,
\begin{align}
\S (\Gamma) &\equiv \int \dx \Bigl( 
\s A^\mu \fdq{\Gamma}{A^\mu} + \s c \fdq{\Gamma}{c} + \s \bar c
\fdq{\Gamma}{\bar c} + \s B \fdq{\Gamma}{B} \nonumber \\
& \qquad\qquad + \fdq{\Gamma}{Y_{\lambda\al}}\fdq{\Gamma}{\lambda^\al} +
\fdq{\Gamma}{Y_{\bar\lambda}^\da}\fdq{\Gamma}{\bar\lambda_\da} \nonumber \\
& \qquad\qquad + \fdq{\Gamma}{Y_{\phi_L}}\fdq{\Gamma}{\phi_L} +
\fdq{\Gamma}{Y_{\phi_L^\dagger}}\fdq{\Gamma}{\phi_L^\dagger} +
  \fdq{\Gamma}{Y_{\psi_L \al}}\fdq{\Gamma}{\psi_L^\al} +
    \fdq{\Gamma}{Y_{\bar\psi_L}^\da} \fdq{\Gamma}{\bar\psi_{L\da}} + (L \to
      R) \Bigr) \nonumber \\
& \quad\quad + \s\eps^\al \pdq{\Gamma}{\eps^\al} + \s \bar \eps_\da
\pdq{\Gamma}{\bar \eps_\da} + \s \omega^\mu \pdq{\Gamma}{\omega^\mu}
 \\
&\equiv \int \left( \s \phi_i' \fdq{\Gamma}{\phi_i'} +
  \fdq{\Gamma}{Y_i}\fdq{\Gamma}{\phi_i} \right)\\
&= 0 \label{STI}
\end{align}
($\phi'$: all linearly transforming field and the ghosts; $\phi$: all
non-linearly transforming fields).
Part of the hypotheses is the gauge fixing as above (this can be
established to all orders) and the ghost dependence (s. \cite{HKS} for
details).

An important calculational tool is given by the linearized ST operator
\begin{align}
\S_\Gamma &\equiv \int \left( \s \phi_i' \fdq{}{\phi_l'} +
  \fdq{\Gamma}{Y_i} \fdq{}{\phi_i} + \fdq{\Gamma}{\phi_i} \fdq{}{Y_i}
  \right)
\end{align}
which satisfies
\begin{equation}
\S_\Gamma \, \S(\Gamma) =0
\label{nilpot}
\end{equation}
provided %(\ref{STI}) holds and
\begin{equation}
\S_\Gamma^2 A_\mu =0\,.
\end{equation}
(The latter relation is true for the final vertex functional.)

As a consequence of gauge fixing, ghost equations and (\ref{STI}) it has
been shown in \cite{HKS} that the following WI's hold
\begin{align}
\pr^\mu\fdq{\Gamma}{A^\mu} &  =  -\i e w_{\rm em}\Gamma -\Box B
 +O(\omega)
\label{gaugeWI}
\ ,\\
w_{\rm em} & =  Q_L\Bigl(\phi_L\fdq{}{\phi_L} -
Y_{\phi_L}\fdq{}{Y_{\phi_L}} + \psi_L\fdq{}{\psi_L}
- Y_{\psi_L}\fdq{}{Y_{\psi_L}} 
\nonumber\\
&\quad - \phi_L^\dagger\fdq{}{\phi_L^\dagger}
 + Y_{\phi_L^\dagger}\fdq{}{Y_{\phi_L^\dagger}}
 - \bar \psi_L\fdq{}{\bar \psi_L}
 - Y_{\bar \psi_L}\fdq{}{Y_{\bar\psi_L}} \Bigr) 
\nonumber\\
&\quad + (L\to R)
\end{align}
and
\begin{equation}
\int \left( \partial_\mu\phi_i' \fdq{\Gamma}{\phi_i'}
           + \partial_\mu\phi_i \fdq{\Gamma}{\phi_i}
           + \partial_\mu Y_i \fdq{\Gamma}{Y_i}\right) =0\,.
\label{translationWI}
\end{equation}
(\ref{gaugeWI}) expresses the local gauge invariance of $\Gamma$, whereas
(\ref{translationWI}) says that $\Gamma$ is translation invariant.

%---------------------------------------------------
\subsection{Susy on the vertex like functionals}
\label{sec:susy-vertex-like} 
%---------------------------------------------------
Like in the Wess-Zumino model (section \ref{sec:supersymm-transf-fun}) we follow
\cite{White, MPW}
and introduce with the help of
the operator 
\begin{equation}
\N \equiv \eps \pdq{}{\eps} + \bar \eps \pdq{}{\bar \eps} + \omega^\mu \pdq{}{\omega^\mu}
\end{equation}
a ``filtration''.
$\S_\Gamma$ will be expanded according to the number of constant ghosts:
\begin{equation}
\S_\Gamma = \sum_{n\ge 0} \S_\Gamma^{(n)}
\end{equation}
where
\begin{equation}
[ \N, \S_\Gamma^{(n)}] = n \S_\Gamma^{(n)}\,.
\end{equation}
We have
\begin{equation}
\S(\Gamma) = \S(\Gamma^{(0)}) + \S_{\Gamma^{(0)}} \Gamma^{(1)} +
O(\epsilon^2)\,. \label{eq:filtGamma}
\end{equation}
Like above the nilpotency (\ref{nilpot}) of $\S_\Gamma$ leads to the
consequence 
\begin{subequations}
\begin{align}
(\S_\Gamma^{(0)})^2 &= 0 \label{eq:SQEDfiltnilpot0}\\
\S_\Gamma^{(0)} \S_\Gamma^{(1)} + \S_\Gamma^{(1)} \S_\Gamma^{(0)} &= 0
\label{eq:SQEDfiltnilpot1} \\
\S_\Gamma^{(0)} \S_\Gamma^{(2)} + \S_\Gamma^{(1)} \S_\Gamma^{(1)}
+ \S_\Gamma^{(2)} \S_\Gamma^{(0)} &=0
\label{eq:SQEDfiltnilpot2}
\end{align}
\end{subequations}
But it is to be noted that the sector with ghost number 0 is now the one of
{\em ordinary} BRS invariance.
Hence $\S_\Gamma^{(0)}$ is the ordinary BRS variation of functionals and
(\ref{eq:SQEDfiltnilpot0}) expresses its nilpotency.

Decomposing $\S_\Gamma^{(1)}$ again according to 
\begin{equation}
\S_\Gamma^{(1)} = \eps^\al W_\al + \bar \eps_\da \bar W^\da + \omega^\mu
W_\mu + \eps \sigma^\nu \bar \eps \fdq{}{\omega^\nu}
\end{equation}
and inserting into (\ref{eq:SQEDfiltnilpot2}) one finds
\begin{align}
\{ W_\al, \bar W_\da \} & \sim 2 \sigma^\mu_{\al\da} W_\mu \\
\{ W_\al, W_\be \} & \sim 0 \sim \{ \bar W_\da, \bar W_\db \}
\end{align}
i.e.\ the supersymmetry algebra on functionals $\F(\phi)$ which obey
\begin{equation}
\S_\Gamma^{(0)} \F =0
\end{equation}
and where $\sim$ means up to $\S_\Gamma^{(0)}$-variations.
On functionals which are not invariant under ordinary BRS transformations
the well-known susy algebra is realized only modulo BRS variations.
Hence one can interprete $W_\al$ and $\bar W_\da$ as giving the
supersymmetry variation of an arbitrary functional.

The simplest ones are the fields themselves and in \cite{HKS} several
examples have been calculated in the one-loop approximation with the
outcome that e.g.\ $\psi(x)$ transforms non-locally under $W_\al$.
The closer analysis of this result is the subject of the subsequent
analysis in terms of general Green functions and then of operators.

%---------------------------------------------------
\subsection{Symmetries on quantum fields}
\label{sec:symm-quant-fields}
%---------------------------------------------------
For the case of space-time translations there is no difference to the
Wess-Zumino model. The WI (\ref{translationWI}) can be turned into a local
one, formulated on the general Green functions Z and and translated into
operator relations analogous to (\ref{eq:transfield}).

Similarly we rewrite the gauge WI (\ref{gaugeWI}) as an eqn. on Z
\begin{align}
-\i \pr^\mu J_\mu^{\rm em} \cdot Z &= -e w_{\rm em}(j) \, Z + \i \Box
 \fdq{Z}{j_B} + o(\omega) \label{WIemloc}\\
0 &= \int \left(  \pr_\mu j_{\phi_i'} \fdq{}{j_{\phi_i'}} + \pr_\mu
  j_{\phi_i}\fdq{}{j_{\phi_i}} + \pr_\mu Y_i \fdq{}{Y_i} \right) Z\,.
\end{align}
and define a conserved electromagnetic current insertion:
\begin{equation}
\left.e w_{\rm em} (j) Z\right|_{Y=0\atop \omega=0} = \i \Box \fdq{Z}{j_B}
+ \pr^\mu J_\mu \cdot Z = \i \pr^\mu j_\mu^{\rm em} \cdot Z
\end{equation}
We find first of all that the corresponding operator is conserved
\begin{equation}
0 = (\pr^\mu j_\mu^{\rm em})^\op\,.
\end{equation}
Differentiating once w.r.t.\ the source $j_{\phi_L}$ and performing LSZ
reduction we obtain
\begin{equation}
-\i e Q_L \delta(y-x) \phi_L^\op(y) = \i \pr^\mu \T \left( j_\mu^{\rm
 em}(y) \phi_L(x)\right)^\op\,.
\end{equation}
Integration yields the transformation for the field operator
\begin{equation}
-\i e Q_L \phi^\op(x) = \i [Q^{\rm em}, \phi_L(x)]^\op
\end{equation}
where
\begin{equation}
Q^{\rm em} \equiv \int \!\!{\rm d}^3y\, j_0^{\rm em}(y)\,,
\end{equation}
similarly for all other field operators.
In particular
\begin{equation}
\i [Q^{\rm em}, A_\mu(x)]^\op =0\,,
\end{equation}
the field $A_\mu$ is not charged \cite{Zimmermann}.

Aiming now at the BRS charge we derive the first consequence from the ST
identity directly. (\ref{STI}) is rewritten on $Z$
\begin{multline}
\S(Z) \equiv \int \dx \Biggl[ 
\i j_{A\mu} \left(\pr_\mu \fdq{Z}{j_c} +\i \eps^\al \sigma_{\mu\al\da}
  \fdq{Z}{\bar \eta_{\lambda \da}} -\i \omega^\nu \pr_\nu
    \fdq{Z}{j_{A\mu}} \right) 
-\i j_c \left( 2\i\eps\sigma^\nu \bar \eps \fdq{Z}{j_A^\nu} -\i \omega^\nu
  \pr_\nu \fdq{Z}{j_c} \right) \\
-\i j_{\bar c} \left(  \fdq{Z}{j_B} -\i \omega^\nu \pr_\nu
  \fdq{Z}{j_{\bar c}} \right)
+ \i j_B \left( 2\i \eps\sigma^\nu \bar\eps \pr_\nu \fdq{Z}{j_{\bar
      c}} -\i \omega^\nu \pr_\nu \fdq{Z}{j_B} \right) 
-\i \eta_\lambda^\al \fdq{Z}{Y_\lambda^\al} -\i \bar \eta_{\lambda \da}
  \fdq{Z}{\bar Y_{\lambda\da}} \\
+\i j_{\phi_L} \fdq{Z}{Y_{\phi_L}} + \i \bar j_{\phi_L} \fdq{Z}{\bar
  Y_{\phi_L}} -\i \eta_{\psi_L}^\al \fdq{Z}{Y_{\psi_L}^\al} -\i \bar
\eta_{\psi_L\da} \fdq{Z}{\bar Y_{\bar\psi_L \da}}
+ (L \longrightarrow R) \Biggr]\\
 -2\i \eps \sigma^\mu \bar \eps \pdq{Z}{\omega^\mu}
 =0
\end{multline}
If a local ST identity is desired we have to permit the ghosts $\eps$,
$\bar \eps$, $\omega^\mu$ to become {\em local}. The local ST identity will
then look as follows
\begin{align}
\S_{\rm loc} Z &\equiv \i j_{A\mu} \left(\pr_\mu \fdq{Z}{j_c} +\i \eps^\al \sigma_{\mu\al\da}
  \fdq{Z}{\bar \eta_{\lambda \da}} -\i \omega^\nu \pr_\nu
    \fdq{Z}{j_{A\mu}} \right) 
-\i j_c \left( 2\i\eps\sigma^\nu \bar \eps \fdq{Z}{j_A^\nu} -\i \omega^\nu
  \pr_\nu \fdq{Z}{j_c} \right) \nonumber \\
& \quad -\i j_{\bar c} \left(  \fdq{Z}{j_B} -\i \omega^\nu \pr_\nu
  \fdq{Z}{j_{\bar c}} \right)
+ \i j_B \left( 2\i \eps\sigma^\nu \bar\eps \pr_\nu \fdq{Z}{j_{\bar
      c}} -\i \omega^\nu \pr_\nu \fdq{Z}{j_B} \right) 
-\i \eta_\lambda^\al \fdq{Z}{Y_\lambda^\al} -\i \bar \eta_{\lambda \da}
  \fdq{Z}{\bar Y_{\lambda\da}} \nonumber  \\
&\quad +\i j_{\phi_L} \fdq{Z}{Y_{\phi_L}} + \i \bar j_{\phi_L} \fdq{Z}{\bar
  Y_{\phi_L}} -\i \eta_{\psi_L}^\al \fdq{Z}{Y_{\psi_L}^\al} -\i \bar
\eta_{\psi_L\da} \fdq{Z}{\bar Y_{\bar\psi_L \da}}
+ (L \longrightarrow R) \nonumber\\
&\quad  -2\i \eps \sigma^\mu \bar \eps \fdq{Z}{\omega^\mu}
\nonumber \\
& =  [\pr^\mu J_\mu +  \pr^\mu \eps^\al K_{\mu\al} + 
\pr^\mu \bar \eps^\da \bar K_{\mu\da} + \pr^\mu \omega^\nu K_{\mu\nu}]
  \cdot Z  
\label{STIlocZ}
\end{align}
Here $J_\mu$ depends on the ghosts $\eps$, $\bar \eps$,
$\omega^\mu$ and for constant ghosts integration will indeed yield zero on
the r.h.s.
At $\eps=\bar\eps=\omega=0$ we are obviously dealing with a local version
of the ordinary ST identity, i.e.\ $\left.\pr^\mu
  J_\mu\right|_{\eps=\bar\eps=\omega=0}$ will correspond to the divergence
of the BRS current.
Reduction leads to the operator equation
\begin{equation}
0 = \left.(\pr^\mu J_\mu)^\op\right|_{\eps=\bar\eps=\omega=0} \equiv
(\pr^\mu J_\mu^{\rm BRS})^\op
\label{BRSconservation}
\end{equation}
The conserved BRS current leads as usual to a time independent charge via
the definition 
\begin{equation}
Q^{\rm BRS} = - \int \dv J_0^{\rm BRS}\,.
\end{equation}
Since (\ref{BRSconservation}) holds on the complete Fock space $Q^{\rm
  BRS}$ commutes with the S-operator on Fock space \cite{Kugo}.

The transformation law for a field operator of type $\phi$
follows from the operator eqn.
\begin{align}
- \delta(y-x) \fdq{\Geff^\op}{Y(y)} &= \i\, \T(\pr^\mu J_\mu(y)
  \phi(x))^\op \Bigr|_{\epsilon=\bar\epsilon=0}
\end{align}
by integrating over space, $\dnx{3}{y}$, and time in the interval $(x_0-\mu,
x_0+\mu)$ with $\mu\to 0$ and leads to
\begin{equation}
\left.\left( \fdq{\Geff}{Y_\phi(x)}
  \right)^\op\right|_{\eps=\bar\eps=\omega=0} = \i [Q^{\rm BRS},
  \phi(x)]^\op\,.
\end{equation}
$\left( \fdq{\Geff}{Y_\phi(x)}\right)^\op$ starts with the non-linear terms
given by the tree approximation and listed in (\ref{BRStrans}); higher order
corrections appear as they contribute to $\Geff$. 

For a linearly transforming field, type $\phi'$, the operator law is the
one of the classical approximation 
\begin{equation}
\s {\phi'}^\op(x) = \i [Q^{\rm BRS}, \phi'(x)]^\op\,.
\end{equation}
The charge $Q^{\rm BRS}$ is nilpotent.

As is clear from the discussion in section \ref{sec:symm-transf-quant}
information on supersymmetry follows from (\ref{STIlocZ}) by
differentiating once with respect to 
a local susy ghost ($\eps^\al(z)$ or analogously $\bar \eps_\da(z)$). 
Doing so and performing LSZ reduction we obtain the operator equation
\begin{equation}
0 = \fdq{}{\eps^\al(z)} \pr^\mu_x J_\mu^\op(x) + \i \, \T \left( \pr^\mu
  J_\mu(x) \, \fdq{\Geff}{\eps^\al(z)} \right)^\op + \pr^\mu_x \delta(x-z)
  \, K_{\mu\al}^\op(x)\,.
\end{equation}
Integration over $z$ yields
\begin{equation}
0 = \pr^\mu_x \, \pr_{\eps^\al} J_\mu^\op(x) + \i \int \dz \T \left( \pr^\mu
  J_\mu(x) \, \fdq{\Geff}{\eps^\al(z)} \right)^\op\,.
\label{susy_cont}
\end{equation}
Hence there is a candidate for a susy current, $\partial_{\eps^\al}
J_\mu^\op(x)$, but it is not conserved. Defining a charge by 
\begin{equation}
Q_\al(t) \equiv - \int \dv \pr_{\eps^\al} J_0^\op(x)
\label{susycharge}
\end{equation}
it will depend on $t$. Integrating (\ref{susy_cont}) over all of $x$-space
leads to
\begin{equation}
0 = \int \dt \pr_t Q_\al(t) - \i \left[ Q^\brs, \pr_{\eps^\al} \Geff
\right]^\op\,.
\end{equation}
Taking the time integral for asymptotic times $t=\pm \infty$ and
identifying there the charges we can write
\begin{equation}
Q_\al^\out-Q_\al^\in = \i \left[ Q^\brs, \pr_{\eps^\al} \Geff
\right]^\op\,.
\label{diffcharge}
\end{equation}
Since $Q_\al^\out$ develops out of $Q_\al^\in$ via the time evolution
operator $S$, the scattering operator,
\begin{equation}
Q_\al^\out = S Q_\al^\in S^\dagger \,,
\end{equation}
(\ref{diffcharge}) implies
\begin{equation}
[Q_\al^\in, S] = -\i [ Q^\brs, \pr_{\eps^\al} \Geff \cdot S]\,.
\label{QS}
\end{equation}
Here we have used that $Q^\brs$ and $S$ commute.
The interpretation of this result is clear: the charge $Q_\al^\in$ which
may be taken to be the generator of susy transformations on the free
in-states does not commute with the $S$-operator, the reason being the
$\eps$-dependence of $\Geff$. Looking at (\ref{Gammagf}) this arises from the
gauge fixing term whose $\epsilon$-dependent part is as a consequence of
the ghost eqn.\ not renormalized.
Matrix elements between physical states however yield a vanishing r.h.s.\
in (\ref{QS}), hence there $Q_\al^\in$ is a conserved charge.

In the derivation of analogous results for field transformations we
concentrate on non-linearly transforming ones. Differentiating (\ref{STIlocZ})
w.r.t.\ $\eps$ and a source for $\phi$ we obtain after LSZ reduction the
operator relation\footnote{The explicit indication $\op$ has been
  suppressed}
\begin{align}
- \delta(y-x)\, \fdq{^2 \Geff}{Y(y)\, \delta \eps^\al(z)} & -\i \, \delta(y-x)
  \T \left( \fdq{\Geff}{Y(y)} \fdq{\Geff}{\eps^\al(z)} \right) \nonumber \\
&= \i \, \T \left( \fdq{}{\eps^\al(z)} \pr^\mu J_\mu(y) \phi(x) \right) - \T
  \left( \pr^\mu J_\mu(y) \fdq{\Geff}{\eps^\al(z)} \phi(x) \right)
  \nonumber  \\
& \quad +\i \, \pr^\mu_y \delta(y-z) \T \left( K_{\mu\al}(y) \phi(x)
  \right)\,.
\label{SUSYop}
\end{align}
Unfortunately one cannot straightforwardly integrate (\ref{SUSYop}) and
conclude how a field $\phi$ transforms under susy because in the T-product
$\T(\pr J \, \cdot\,\pr_\epsilon \Geff \,\cdot\,  \phi)$ distributional singularities at
coninciding points may arise.
It is clear that every renormalization scheme leads to a well-defined
T-product which is integrable in the sense of distributions but of which
type this distribution is has to be found out.
We present this analysis in the appendix and quote here only its result.
One finds that contributions with $\delta(y-x)$ arise which  cancel
with the operator product $\T( 
\delta\Geff / \delta Y(y) \,\cdot\, \delta \Geff / \delta\epsilon^\al(z))$ on
the l.h.s., and in addition a term with a double delta function
$\delta(x-y)\delta(y-z)$ which contributes to the susy variation of
the photino field $\lambda$ only.
Altogether we arrive at
\begin{align}
\fdq{}{Y_\phi(x)}\pr_{\eps^\al} \Geff^\op &= \i \left[ Q_\al(x^0),
  \phi(x) \right]^\op \,, \qquad \text{if } \phi \ne \lambda \\
\fdq{}{Y_\lambda^\be(x)}\pr_{\eps^\al} \Geff^\op  + \eps_{\al\be} B^\op(x)
  &= \i \left[ Q_\al(x^0),
  \lambda_\be(x) \right]^\op \,.
\label{eq:lambdatransfinal}
\end{align}

Hence the time-dependent susy charge defined in (\ref{susycharge})
$(t\equiv x_0)$ generates a local, non-linear
transformation on all fields.
The fact that this transformation does not correspond to a symmetry on the
Fock space, but only on the Hilbert space of the theory is completely
encoded in the time dependence of the charge $Q_\al(t)$ and in the
additional $\eps_{\al\be} B$ term in the susy transformation of
$\lambda^\al$ which vanishes between physical states.

Comparing this result with the explicit one-loop calculation in \cite{HKS}
of $\S_\Gamma \psi(x)$ which yielded a non-local expression we can trace
the origin of this non-locality: it comes from the operator product
$\delta\Geff/\delta Y(y) \, \cdot \, \delta\Geff/\delta\epsilon^\al(z)$ in
the l.h.s.\ of (\ref{SUSYop}) which could be separated in the above
analysis and seen to cancel against the $\delta(y-x)$ terms on the r.h.s.
By going to the operator level one could isolate the local
transformation. The transformation on vertex type functionals given by
$\S_\Gamma^{(1)}$ does not distinguish between these different
contributions since it fixes in a sense susy transformations only ``up to''
BRS variations.

The somewhat surprising modification (\ref{eq:lambdatransfinal}) of
the transformation of 
$\lambda^\al$ may be understood as follows.
We have defined a time-dependent susy charge $Q_\al(x^0)$ which
generates susy transformations $[Q_\al(x^0), \lambda_\be(x^0, \vec x)]$
at time $x^0$. The time dependence of all these operators is
determined by the same unitary time evolution operator in Fock
space, i.e.\ our susy transformation is compatible with this time
dependence. However, the equation of motion for $\lambda^\al$ contains
contributions from the gauge fixing term which are not supersymmetric
and therefore the time dependence of $\lambda^\al$ cannot be
compatible with ordinary susy. This shows that $Q_\al(x^0)$ cannot generate the
standard susy transformation of $\lambda^\al$.
More explicitly, the equation of motion for $\bar \lambda^\da$ has the
form
\begin{equation}
\i \sigma^\mu_{\al\da} \pr_\mu \lambda^\al = \bar F_\da  + \i \eps^\al
\sigma^\mu_{\al\da} \pr_\mu
\bar c \,,
\end{equation}
where $\bar F_\da$ contains all interaction terms of the classical action
(\ref{eq:GammaSQED}). This equation is invariant under the combined
gauge+susy BRS 
transformations (\ref{BRStrans}), which means to first order in $\eps$:
\begin{equation}
\i \sigma^\mu_{\al\da} \pr_\mu \left( \s^{\rm susy}
  \lambda^\al\right)  = \s^{\rm susy} \bar F_\da  + \i \eps^\al
  \sigma^\mu_{\al\da} \pr_\mu \left( \s^{\rm gauge}
  \bar c \right) \,,
\label{eqmosusy}
\end{equation}
where
\begin{align}
\s^{\rm gauge} &= \left. \left( \eps^\al \pdq{}{\eps^\al} + \bar
    \eps^\da \pdq{}{\bar\eps^\da} \right) \s
    \right|_{\eps=\bar\eps=\omega=0} \\ 
\s^{\rm susy} &= \left. \s \right|_{\eps=\bar\eps=\omega=0} \,.
\end{align}
Like in the filtration (\ref{eq:filtGamma}), supersymmetry holds only
up to gauge-BRS transformations, in particular we have
\begin{equation}
\i \sigma^\mu_{\al\da} \pr_\mu \left( \s^{\rm susy}\lambda^\al\right)
\ne \s^{\rm susy} \bar F_\da  \,.
\end{equation}
Since $s^{\rm gauge} \bar c = B$, (\ref{eqmosusy}) may be rewritten as
\begin{equation}
\i \sigma^\mu_{\al\da} \pr_\mu \left( \s^{\rm susy} \lambda^\al -
 \eps^\al B \right) = \s^{\rm susy} \bar F_\da  \,.
\end{equation}
This shows that the modification of the susy transformation law of
$\lambda^\al$  ensures compatibility with time evolution.

%---------------------------------------------------------------
\section{Discussion and Conclusions}
\label{sec:disc-concl}
\setcounter{equation}{0}

%---------------------------------------------------------------
Comparing the results of our analysis for the Wess-Zumino model and SQED in
the Wess-Zumino gauge we find
\begin{itemize}
\item Closely analogous ones: translations and gauge transformations obey
  simple WIs and can easily be interpreted in terms of operators; susy
  requires the use of the generalized ST identity expressing the
  generalized BRS invariance; its non-linear character does not really
  cause harm;
\item clearly different ones: in the Wess-Zumino model susy is not broken
  although non-linearly realized; its charge exists and is time
  independent, the transformation is local; in SQED susy is broken by the
  gauge fixing term and the BRS type formulation only enables one to carry
  that breaking along in a fashion which permits renormalization and
  consistent treatment to all orders (the gauge fixing term is a BRS
  variation); the susy charge is time dependent, but in such a way that the
  dependence disappears between physical states; remarkably enough there
  exists still a local operator expression for the field
  transformations.
  The susy transformation of the field $\lambda^\al$ is modified by
  the term $\eps_\al B$ which vanishes between physical states.
\end{itemize}
The comparison with the linear realization of supersymmetry which is
possible in these two examples yields qualitative agreement. In the
Wess-Zumino model the respective charges ($P_\mu$, $Q_\al$, $\bar Q_\da$)
exist and operate on a Hilbert space which is essentially the same as the
one generated without auxiliary fields. The latter can be understood as
interpolating fields which are useful but not necessary.

In the massive SQED the charges $P_\mu$, $Q_\al$, $\bar Q_\da$ also exist,
likewise the ``electric'' charges because there exist conserved currents
which are gauge invariant. All transformations are linear, given by WIs
hence the implementation on the Fock space is straightforward. The
transformation laws for operators are local. Here, of course, the Fock
spaces differ tremendously, but one expects (although a detailed proof
seems to be missing) that the Hilbert spaces are equivalent and on them the
theories should coincide,  in particular the charges $P_\mu$, $Q_\al$,
$\bar Q_\da$ and the (true) electric charge $Q$.

A first comment is in order as far as our rather careless treatment of
operators in SQED with massless photon and photino is concerned. 
With some additional equipment we could have introduced also in the
Wess-Zumino gauge a mass term for photon and photino (namely by
admitting some susy partners for $c$, $\bar c$ etc.). 
With an eye on the linear realization we did not do so: our main
interest was to clarify the role of the gauge fixing term and the
understanding of the breaking of susy which its presence causes in the
Wess-Zumino gauge formulation of supersymmetry. 
The infraredwise existence of operators was not our concern -- in this
respect our work is formal.

A second comment is appropriate as far as the breaking of susy is
concerned. Since it takes place only in Fock space but not in Hilbert space
we have the impression that the ``collapse of supersymmetry'' as described
in the paper by Buchholz, Ojima et al.\ (s. \cite{collapse}) is still
avoided for the physical part of the theory.

A third comment finally refers to the type of operator equations which
we have derived. Obviously we worked in perturbation theory and
therefore in a very specific representation of the (anti-) commutation
relations. Hence these equations cannot be considered as abstract
ones, valid for any arbitrary representation\footnote{We are very much
  indebted to H.\ Grosse for clarifying discussions of this point.}.
In fact, it seems to be an open problem in which sense they could be
``lifted'' to have representation independent meaning.
Or stated differently: it is not known which equations valid in
perturbation theory hold also beyond perturbation theory.
On the other hand it is also obvious that the abstract
non-perturbative approach to quantum field theory is not yet able to
handle the type of charges dealt with in this paper.

\begin{appendix}
%---------------------------------------------------------------
\section{Singularity structure of some T-products}
\label{sec:appendix}
\setcounter{equation}{0}

%---------------------------------------------------------------

The starting point for the present analysis is the local ST identity
\begin{align}
\S_{\rm loc} Z &= \left( \pr^\mu J_\mu + \pr^\mu \epsilon^\al K_{\mu\al} +
  \pr^\mu \bar \epsilon^\da  K_{\mu \da} + \pr^\mu\omega^\nu  K_{\mu\nu} 
   \right) \cdot Z
\end{align}
with with $\S_{\rm loc}$ given by (\ref{STIlocZ}).
Since we are mainly interested in the properties of a non-linearly
transforming field, say $\phi(x)$, under superymmetry we differentiate
w.r.t.\ the source for $\phi$ and $\epsilon^\al(z)$. This gives for a
general Green function
\begin{multline}
\delta(y-x) 
\left( \left\langle \fdq{^2 \Geff}{\epsilon^\al(z) \, \delta Y_{\phi}(x)} \,
      X \right\rangle + \left\langle \fdq{\Geff}{\epsilon^\al(z)} \,
      \fdq{\Geff}{Y_\phi(x)} \, X \right\rangle \right) \\
  \shoveleft{+ \sum_k \delta(y-x_k) \left( \left\langle \fdq{^2
          \Geff}{\epsilon^\al(z) 
          \, \delta Y_{\phi_k}(x_k)} \,\phi(x)\, 
        X_{\check k} \right\rangle 
      + \left\langle \fdq{\Geff}{\epsilon^\al(z)} \,
        \fdq{\Geff}{Y_{\phi_k}(x_k)} \, X_{\check k} \right\rangle \right)} \\
  \shoveleft{ + \sum_{k'} \delta(y-x_{k'}) \left( \left\langle 
        \left(\fdq{(\s \phi_{k'}(y))}{\epsilon^\al(z)} \right)\, \phi(x)
        \, X_{\check k'} \right\rangle 
      + \left\langle \s \phi_{k'}(y) \, \fdq{\Geff}{\epsilon^\al(z)} \,
        \phi(x) \, X_{\check k'} \right\rangle \right) } \\
  \shoveright{= \left\langle \left( \fdq{\pr^\mu
          J_\mu(y)}{\epsilon^\al(z)}\Bigr|_{\epsilon=0} + \delta(y-z) \pr^\mu
        J_{\mu\al}(y) \right) \, \phi(x) \, X \right\rangle} \\
  + \left\langle \pr^\mu J_\mu(y) \Bigr|_{\epsilon=0} \,
    \fdq{\Geff}{\epsilon^\al(z)} \, \phi(x) \, X \right \rangle
\label{eq:A3}
\end{multline}
Here denotes $X$ a general string of fields, $X_{\check k}$ indicates that
the field with index $k$ is missing, the same for $X_{\check k'}$. Fields
$\phi'$ transform linearly. $\pr^\mu J_\mu|_{\epsilon=0}=\pr^\mu
J_\mu^{\rm BRS}$ is the conserved BRS current.

The problem is now to identify within
\begin{align}
\delta(y-x) \left( \fdq{^2\Geff}{\epsilon(z)\, \delta Y_\phi(y)} +
  \T \left( \fdq{\Geff}{\epsilon(z)} \, \fdq{\Geff}{Y_\phi(y)} \right)
  \right) 
& = \T \left(
  \fdq{\pr^\mu J_\mu(y)}{\epsilon(z)} \, \phi(x)
  \right) \nonumber \\
& \quad +  \pr_y^\mu \delta(y-z) \T \left( 
  K_{\mu\al}(y)\,  \phi(x)
  \right) \nonumber \\
& \quad + \T \left( \pr^\mu J_\mu^{\rm BRS}(y) \,\,
  \fdq{\Geff}{\epsilon(z)} \,\,  \phi(x) \right)
\label{eq:A4}
\end{align}
the singularities which form $\delta(y-z)$ or $\delta(z-x)$ or both. These
singularities determine the possible contributions from the T-product when
it is integrated over $y$ and $z$.
In particular, we would like to integrate in the order
\begin{equation}
\int\limits_{x_0-\mu}^{x_0+\mu} \dnx{}{y_0} \int \dnx{4}{z} \int
\dnx{3}{y}
\label{eq:intorder}
\end{equation}
and take the limit $\mu\to 0$. Clearly, we can get a contribution only
from terms containing a factor $\delta (x-y)$. In the triple insertion
\begin{equation}
\left\langle \pr^\mu J_\mu^\BRS(y) \,
    \fdq{\Geff}{\epsilon^\al(z)} \, \phi(x) \, X \right \rangle\,,
\label{tripleins}
\end{equation}
$\delta$-functions may arise when a suitable number of partial
derivatives form a wave equation operator and act on the corresponding
propagator, e.g.
\begin{align}
  \Box \langle c\, \bar c \rangle (y-z) &= \i \delta(y-z)\,, 
 \label{ccprop}\\
  \Box \langle B \, A_\mu \rangle (y-u) &= \i \pr_\mu
  \delta(y-u)\,, \label{BAprop} \\
  \sigma^\mu_{\al\da} \pr_\mu \langle \lambda_\be \,  \bar
  \lambda^\da \rangle (y-z) &= \eps_{\al\be}
  \delta(y-z)\,. \label{llprop} 
\end{align}
In terms of diagrams, the propagator shrinks then to a
point and is replaced by a $\delta$-function.

We first consider the case that $\pr^\mu J_\mu^\BRS(y)$ and $\phi(x)$
are directly connected by a line, leading to a contribution with
$\delta(x-y)$. 
Comparing with the WI for ordinary BRS transformations into which one
inserts a vertex $\fdq{\Geff}{\epsilon}$ ``by hand'' it is clear
that such  contributions  cancel
against the  double insertion
\begin{equation}
\delta(y-x) \left\langle  \fdq{\Geff}{\epsilon} \,
  \fdq{\Geff}{Y_\phi}\, X\right\rangle\,
 \label{doubleins}
\end{equation}
on the l.h.s.\ of (\ref{eq:A3}),
including all possible double delta functions
$\delta(y-x)\delta(y-z)$.

However, there is a second possibility how $\delta(x-y)$ may be
generated: If the vertices are connected in the order
\[
 \phi(x) --- \fdq{\Geff}{\eps(z)} --- \pr^\mu J_\mu^\BRS(y)
\]
we may obtain a double $\delta$ function $\delta(x-z)\delta(z-y)$
which is equivalent to  $\delta(x-y)\delta(z-y)$ but is not present in
the double insertion (\ref{doubleins}).
This can only happen for $\phi=\lambda^\al$ (or $\bar \lambda^\da$), the relevant
contributions being
\begin{align}
\pr^\mu J_\mu^{\rm BRS} &= B \Box c + \dots \\
\fdq{\Geff}{\eps^\al} &= \i \bar c \sigma^\mu_{\al\da} \pr_\mu \bar
\lambda^\da + \dots\,.
\end{align}
\begin{equation}
\raisebox{-1.5cm}{\includegraphics[bb=104 626 248 724]{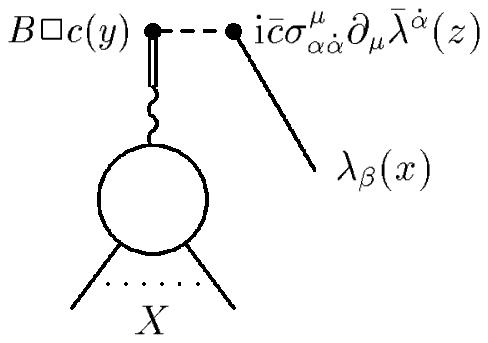}}
\longrightarrow\quad \delta(y-z) \,\delta(x-z)\hspace*{-4mm}
\raisebox{-1.5cm}{\includegraphics[bb=104 626 182 724]{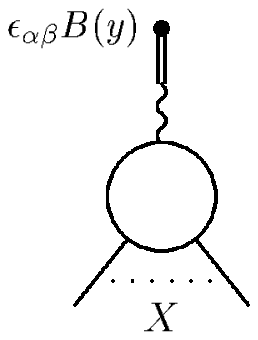}}
\label{eq:diagram}
\end{equation}
Here, the lines \includegraphics[bb=125 705 152 713]{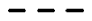} and
 \includegraphics[bb=125 705 161 713]{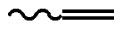} represent propagators
 $\langle c \, \bar c \rangle$ resp.\ $\langle A^\nu\, B\rangle$.

According to (\ref{ccprop}), (\ref{llprop}), the diagram on the
l.h.s.\ of (\ref{eq:diagram}) corresponds
to a contribution  
\begin{equation}
\delta(x-y)\delta(y-z) \langle \eps_\al B(x) \, X \rangle\,.
\label{Bcontribution}
\end{equation}
We will now further investigate this contribution in order to clarify
for which field configurations $X$ it is really non-vanishing.
From the Ward identity (\ref{gaugeWI}), it follows immediately that the
operator $B^\op$ is a free field,
\begin{equation}
\Box B^\op =0\,.
\end{equation}
Therefore there is only one contribution to the
diagram on the right hand side of (\ref{eq:diagram}), namely
\begin{equation}
\raisebox{-1.7cm}{\includegraphics[bb=124 616 182 724]{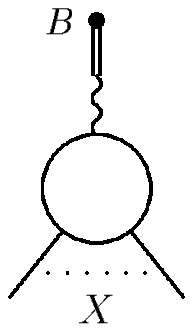}}
\quad=\quad
\raisebox{-1.7cm}{\includegraphics[bb=104 616 196 724]{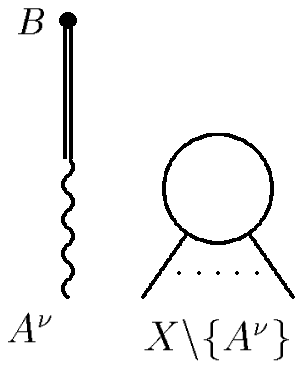}}
\quad+ \mbox{contact terms}
\end{equation}
Reduction w.r.t.\ $A^\nu$ yields zero, but reduction w.r.t.\ $B$ is
non-zero. Thus we see that the insertion
of $\eps_\al B$ (\ref{Bcontribution}) is indeed nonvanishing, but it
is made up 
entirely of trivial contributions.

Taking into account the above
discussions, (\ref{eq:A4}) reads
\begin{align}
\delta(y-x) \fdq{^2 \Geff}{\eps \delta Y_\phi} &= \T \left( \left(
    \fdq{\pr^\mu J_\mu}{\eps} + \pr^\mu \delta(y-z) K_{\mu\al} \right) \, \phi(x)
    \right) + \text{less singular} \,, \qquad \text{if }\phi \ne
    \lambda^\al \\
\delta(y-x) \fdq{^2 \Geff}{\eps \delta Y_\lambda^\al } &= \T \left( \left(
    \fdq{\pr^\mu J_\mu}{\eps} + \pr^\mu\delta(y-z)  K_{\mu\al} \right) \, \lambda^\al(x)
    \right) + \delta(y-x) \delta(y-z) \epsilon^\al B(x) \nonumber \\
 & \quad + \text{less singular}\,.
\end{align}
Here, ``less singular'' stands for terms containing no singularity of
the type $\delta(x-y)$.
Integration in the order (\ref{eq:intorder}) yields a local susy
variation for all fields,
\begin{align}
\fdq{}{Y_\phi(x)}\pr_{\eps^\al} \Geff^\op &= \i \left[ Q_\al(x^0),
  \phi(x) \right]^\op \,, \qquad \text{if } \phi \ne \lambda \\
\fdq{}{Y_\lambda^\be(x)}\pr_{\eps^\al} \Geff^\op  + \eps_{\al\be} B(x)
  &= \i \left[ Q_\al(x^0),
  \lambda_\be(x) \right]^\op \,.
\end{align}

We still note that these expressions are covariant, i.e.\ renormalization
scheme independent, because they are unique: after the double insertion
contributions cancelled each other the covariance is guaranteed.

\end{appendix}

\noindent{\bf Acknowledgement}\\
We thank Harald Grosse for helpful discussions.


\begin{thebibliography}{99}
\bibitem{GSW}S.~L.~Glashow,
{\sl Partial Symmetries Of Weak Interactions},
Nucl.\ Phys.\  {\bf 22} (1961) 579.\\
S.~Weinberg,
{\sl A Model Of Leptons},
Phys.\ Rev.\ Lett.\  {\bf 19} (1967) 1264.\\
A.~Salam, in: {\sl Proceedings of the 8th Nobel Symposium}, p. 367,
ed. N. Svartholm, Almquist and Wiksell, Stockholm 1968.
\bibitem{elweak}W.~Hollik and G.~Duckeck,
``Electroweak precision tests at LEP,'' {\it  Berlin, Germany: Springer
  (2000) 161 p}. \\
M.~W.~Grunewald,
``Experimental tests of the electroweak standard model at high energies,''
Phys.\ Rept.\  {\bf 322} (1999) 125.
\bibitem{EKhabil}E.~Kraus,
   {\sl Renormalization of the electroweak standard model to all orders},
   Annals Phys.\  {\bf 262} (1998) 155,
   hep-th/9709154.
\bibitem{HKS}
   W.~Hollik, E.~Kraus, D.~St{\"o}ckinger,
   {\sl Renormalization and symmetry conditions in supersymmetric QED},
   Eur.\ Phys.\ J.\  {\bf C11} (1999) 365, hep-ph/9907393.
\bibitem{White}P.L.~White, {\sl An analysis of the cohomology
    structure of super Yang-Mills coupled to matter}, Class.\ Quantum
  Grav.\ {\bf 9} (1992)   1663,\\
P.L.~White, {\sl Analysis of the superconformal cohomology structure
  of $N=4$ super Yang-Mills}, Class.\ Quantum Grav.\ {\bf 9} (1992) 413\,.
\bibitem{MPW}
   N.~Maggiore, O.~Piguet and S.~Wolf,
   {\sl Algebraic renormalization of $N=1$ supersymmetric gauge theories},
   Nucl.\ Phys.\  {\bf B458} (1996) 403,
   hep-th/9507045.
\bibitem{WZgauge}J.~Wess, B.~Zumino,
{\sl Supergauge Invariant Extension Of Quantum Electrodynamics},
Nucl.\ Phys.\  {\bf B 78} (1974) 1.
\bibitem{PSWZ}O.~Piguet, K.~Sibold, {\sl Renormalizing Supersymmetry
    without Auxiliary Fields}, Nucl.~Phys. {\bf B 253} (1985) 269.
\bibitem{Zimmermann}W.~Zimmermann, Lectures on Electrodynamics 1978, unpublished.
\bibitem{Kugo}T.~Kugo, {\sl Eichtheorie}, Springer-Verlag Berlin
  Heidelberg New York 1997.
\bibitem{collapse}D.~Buchholz, I.~Ojima,
{\sl Spontaneous collapse of supersymmetry},
Nucl.\ Phys.\  {\bf B498} (1997) 228,
hep-th/9701005.


\end{thebibliography}
\end{document}